\documentclass[aps,prl,twocolumn,superscriptaddress,showpacs]{revtex4-1}
\pdfoutput=1

\usepackage{graphicx}
\usepackage{amsmath,amssymb}

\newcommand{\ket}[1]{\left| #1 \right\rangle}
\newcommand{\op}[1]{{\bf #1}}
\newcommand{\eins}{\mathbbm{1}}
\renewcommand{\eins}{\mathrm{1}}
\renewcommand{\exp}[1]{\textup{e}^{#1}}
\renewcommand{\i}{\textup{i}}

\begin{document}

\title{Towards $T_1$--limited magnetic resonance imaging using Rabi beats}
\date{2010-09-03}

\author{H. Fedder}\email{helmut.fedder@gmail.com}
\author{F. Dolde}
\author{F. Rempp}
\author{T. Wolf}
\affiliation{3\textsuperscript{rd} Physics Institute and Research
Center SCoPE, University of Stuttgart, Pfaffenwaldring 57, 70550 Stuttgart}

\author{P. Hemmer}
\affiliation{Department of Electrical and Computer Engineering, Texas A\&M
University, College Station, TX 77843, USA}

\author{F. Jelezko}
\author{J. Wrachtrup}
\affiliation{3\textsuperscript{rd} Physics Institute and Research
Center SCoPE, University of Stuttgart, Pfaffenwaldring 57, 70550 Stuttgart}

\begin{abstract}
Two proof-of-principle experiments towards $T_1$-limited magnetic resonance
imaging with NV centers in diamond are demonstrated. First, a large number of
Rabi oscillations is measured and it is demonstrated that the hyperfine
interaction due to the NV's $^{14}$N can be extracted from the beating
oscillations. Second, the Rabi beats under V-type microwave excitation of the
three hyperfine manifolds is studied experimentally and described theoretically.
\end{abstract}

\maketitle

\section{Introduction}

The ever increasing need for high resolution imaging in the life sciences,
material science, and more recently quantum information processing, has led over
the past decade to the development of a variety of methods that try to overcome
the limits set by optical diffraction \cite{pertsinidis2010subnanometre}.
Exploiting the non-linear behaviour of fluorescent labels, a spatial resolution down to few nanometers was
demonstrated using optical techniques such as STED \cite{hell1994breaking}, PALM
\cite{betzig2006science-PALM}, and STORM \cite{rust2006natmeth-STORM}. A
technique whose resolution is independent of the wavelength, is magnetic resonance imaging. Magnetic resonance imaging --and nano scale magnetic-
and electric field sensing-- are key technologies in various areas of science.
Recently, magnetic resonance imaging as well as scanning probe magnetic imaging
on scales relevant to molecular biological processes was demonstrated
\cite{balasubramanian2008nanoscale,maze2008nanoscale}. A spin label that is
suitable for high resolution measurements are nitrogen vacancy centers embedded
in diamond nano crystals \cite{tisler2009fluorescence}. Owing to their  optical
adressability, photostability and long spin coherence time recently sub 10nm spatial resolution as well as the detection of magnetic
fields close to individual electron spins has been demonstrated using NV
nano sensors\cite{balasubramanian2008nanoscale,maze2008nanoscale}. The spatial
resolution of a magnetic resonance measurement is determined by the strength of the field gradient and the effective linewidth
of the electron spin transition. Pulsed
imaging modes, such as the Hahn-Echo and CPMG sequence can be
used to decrease the effective linewidth. The continuous driving of Rabi
oscillations has been proposed to decrease the effective linewidth down to the
limit given by the spin decay time \cite{shin2009sub}. In this paper we perform
a proof-of-principle study of this imaging mode. The feasibility of the method
is demonstrated by resolving the hyperfine interaction of the NV center due to
its $^{14}N$ nuclear spin.

In the simplest magnetic resonance imaging mode, a CW electron spin resonance
(ESR) measurement is performed, and the effective linewidth is given by the spin
dephasing time $T_2^*$, which is typically short and only of the order of few
micro seconds in case of the NV center. In the Hahn-Echo sequence,
$\pi/2-\pi-\pi/2$ pulses are used to refocus precessing spins, such that noise
generated by slowly fluctuating spins in the environment is canceled. For the
Hahn-Echo sequence, the effective linewidth is given by the dephasing time $T_2$
of the spin system, which is larger than $T_2^*$, and can reach several hundred
microseconds in the  case of the NV center \cite{balasubramanian2009ultralong}.
More advanced pulse sequences, such as the CPMG sequence \cite{meiboom1958modified}, hold promise to
decrease the effective linewidth down to the limit set by $T_1^\rho$
\cite{PrinciplesOfPulsedElectronParamagneticResonance,deLange2010single}.
Another elegant imaging mode is based on the continuous driving of Rabi oscillations. The decay time of Rabi oscillations is given precisely by $T_1^\rho$ and depends on
the microwave power. When the microwave power is high and the Rabi frequency
approaches the transition frequency -- corresponding to the limit of the
rotating wave approximation-- $T_1^\rho$ approaches $T_1$. Among the various  magnetic
resonance imaging modes, this mode may potentially achieve the longest decay
time and highest resolution. Over the past years there has been a continued
effort to apply this magnetic resonance imaging mode to the NV center \cite{shin2009sub}. However, this poses
great challenges. In this mode, spatial information is extracted from beat
frequencies observed in the Rabi oscillations. A high spatial
resolution requires the measurement of a large number of Rabi oscillations. Here
we demonstrate the feasibility of this imaging mode by realising the
measurement of a large number ($>500$) of Rabi oscillations, and resolve the
hyperfine splitting due to the NV's $^{14}N$ nuclear spin
from the Rabi beats. Moreover, we investigate experimentally and describe
theoretically the observed Rabi beats in case of a V-type energy level scheme,
which is the desired excitation mode in a high resolution experiment.

\begin{figure}
  \centering
  \includegraphics[scale=0.2]{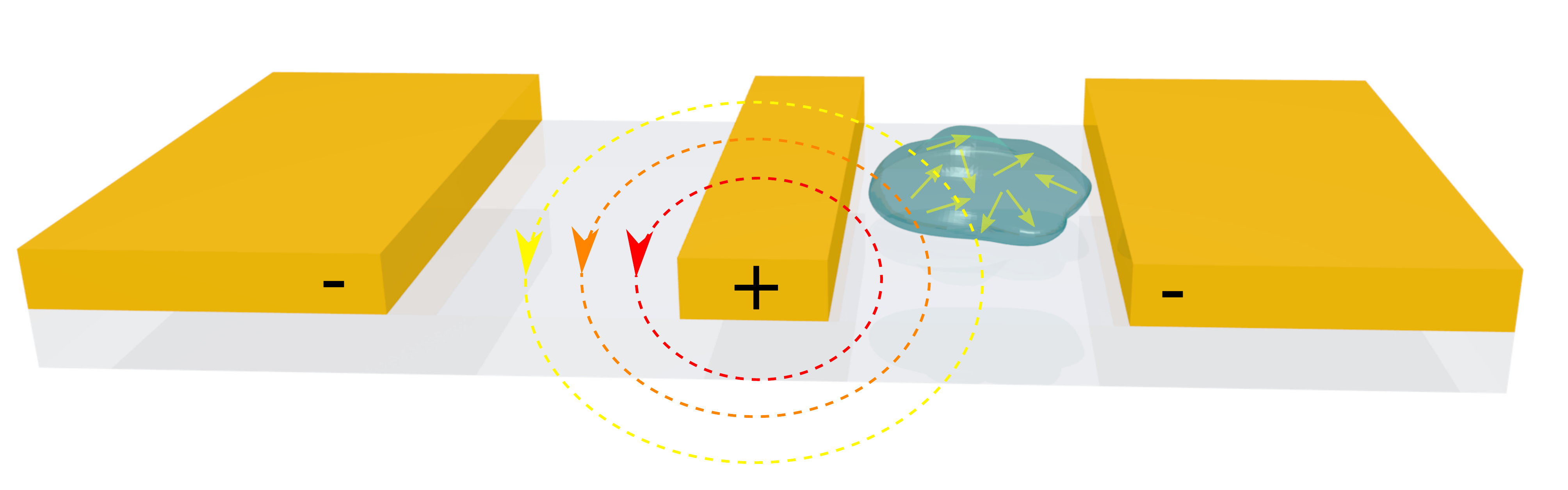}
  \caption{Principle of magnetic resonance imaging based on Rabi beats.}
  \label{fig:principle}
\end{figure}

The principle of magnetic resonance imaging using Rabi beats is illustrated in
Fig.~\ref{fig:principle}. We envision a sample, such as a living cell, that
is tagged with nano diamonds that contain single NV-centers. The NV electron spins are
inititalized and read out optically using a confocal microscope. An
inhomogeneous microwave field -- such as the one generated by a coplanar
waveguide -- is applied to the sample. The microwaves are matched to the
$m_s=0\rightarrow \pm 1$ electron spin transition and drive Rabi
oscillations between the groundstate spin sublevels. The Rabi frequency depends
on the microwave power, thus encoding the position and angle within the
microwave field. From an a priori knowledge of the microwave field
(or from a reference measurement), the position and angle of the individual
NV-centers with respect to the electrodes can be determined. To obtain the NV position in
two spatial directions, two successive measurements are performed with microwave 
gradients applied in different directions using a two dimensional waveguide
structure.

In this magnetic resonance imaging mode, the resolution is given by the decay
time $T_1^\rho$ of the Rabi oscillations, the Rabi frequency $\Omega$, and the
strength of the microwave gradient. The number $N$ of observable Rabi
oscillations is linked to $T_1^\rho$ and the Rabi frequency as
\begin{equation}
\frac{\Omega T_1^\rho}{2\pi}=N,
\end{equation}\label{eq:N}
For the present coplanar waveguide, the
field gradient is determined by the width $G$ of the gap, and the resolution
$\delta x$ follows approximately as
\begin{equation}
\delta x=\frac{G}{N}.
\end{equation}
The maximum achievable Rabi frequency is of the order of the Lamor
frequency of the spin \cite{chiorescu2003coherent}. In this case, the decay time
$T_1^\rho$ of the Rabi signal becomes $T_1$ and the resolution approaches
\begin{equation}
\delta x=\frac{2\pi G}{T_1 \Omega},
\end{equation}
where $\Omega=\omega=D\approx 2.88$GHz is equal to the electron spin
transition. For NV centers, the $T_1$ time is typically few ms at room
temperature \cite{redman1991spin}, such that a ratio $T_1/\omega=10^{-6}$ could
in principle be achievable. With $G=10\mu$m a spatial resolution of $0.01$nm would be
reached. However, this poses stringent requirements on the stability of the
applied microwave field. Both the power stability and the spatial stability of the microwave field must be at least as
good as the ratio $T_1^\rho/\Omega$, and to achieve sub Angstrom resolution,
the microwave wire should not drift by more than a fraction of an atomic 
diameter.

\section{Results and Discussion}

\begin{figure}
  \centering
  \includegraphics{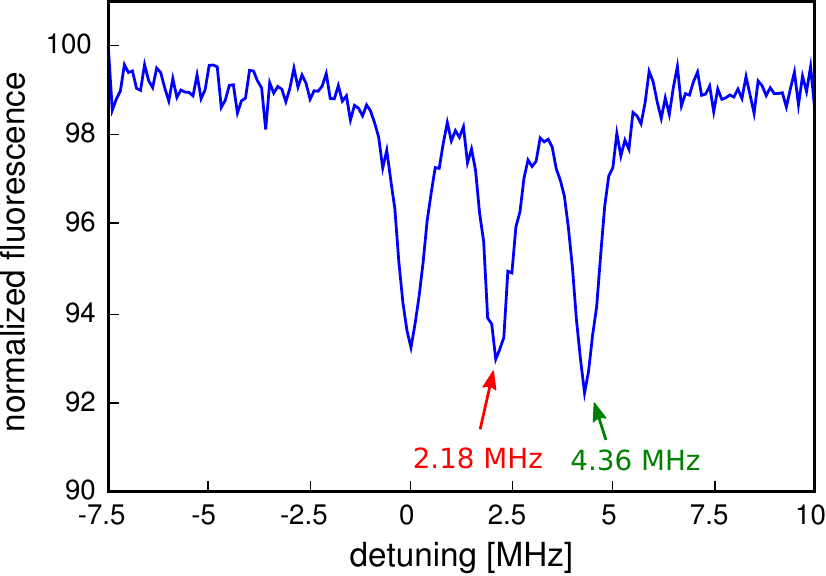}
  \caption{ESR spectrum of the NV center showing the hyperfine splitting
  of the $m_s=0\rightarrow -1$ spin level due to $^{14}N$. The spectrum was
  recorded with an applied DC magnetic field of about 150G. Detuning is denoted
  relative to the lowest hyperfine transition corresponding to 2.4524GHz.}
  \label{fig:esr}
\end{figure}

To demonstrate the observation of Rabi beats, we study a single NV-center in a
fixed microwave field. Rabi beat measurements with a single NV center are
possible owing to hyperfine interaction with the NV's $^{14}N$ nuclear spin, that
results in a splitting of the groundstate spin manifold into several hyperfine
sublevels. Each hyperfine transition has a slightly different Rabi frequency,
which results in a beat signal in the measured Rabi oscillations. These hyperfine
beats will also be seen in high resolution Rabi beat imaging data as a
modulation, and it is imortant to identify and assign them correctly. In here we
are interested in the hyperfine interaction caused by the $^{14}N$ nucleus
(nuclear spin $J=1$) of the NV center, which splits the $m_s=\pm 1$ ground state
spin levels each into three hyperfine sublevels with an energy splitting of
$\delta\approx 2.18$MHz. This energy splitting is seen in the CW electron spin
resonance (ESR) measurement shown in Fig.\ref{fig:esr}. In here, the
ESR measurement is performed on the $m_s=0\rightarrow -1$ electron spin
transition. Zero detuning corresponds to a driving microwave field of 2.4524GHz.
Note, that a permanent magnetic field with a component along the NV axis of about
150G has been applied to shift the $m_s=0\rightarrow -1$ transition away from its
zero field value (2.88GHz) owing to the Zeeman effect, and thus to separate it
from the $m_s=0\rightarrow +1$ spin transition (which is shifted to higher
frequency of 3.3GHz). This separation between the two electron spin transitions
is necessary in order to ensure that the strong  microwave field does not
simultaneously drive both spin manifolds.

We now consider the Rabi beats expected for the given hyperfine splitting.
For a driving microwave field that is detuned by a small frequency $\delta$
from the resonant transition, the Rabi frequency $\Omega$ is increased as
compared to the resonant Rabi frequency $\Omega_0$, following
\cite{scully01quantum_optics}
\begin{equation}
\Omega=\sqrt{\Omega_0^2+\delta^2}.
\end{equation}\label{eq:rabi}
For small detuning ($\delta\ll\Omega_0$), the shift of the Rabi frequency
$\delta\Omega$ becomes
\begin{equation}
\delta\Omega=\frac{\delta^2}{2\Omega_0}.
\end{equation}\label{eq:detuning}
Suppose the microwave frequency corresponds to the lowest hyperfine transition
(as in the present experiments). In this case, the lowest hyperfine transition
is driven with Rabi frequency $\Omega_0$, while the two other hyperfine 
transitions are driven with faster Rabi frequencies corresponding to detunings 
$\delta_0=2.18$MHz and $\delta_{-1}=4.36$MHz. Using Eq.\ref{eq:rabi} we can
evaluate the expected Rabi frequencies. The resulting signal is the sum of
three oscillations, whose frequencies can be extracted by Fourier
transformation. Also, the three frequencies result in several beat
frequencies in the Rabi signal.

\begin{figure}
  \centering
  \includegraphics{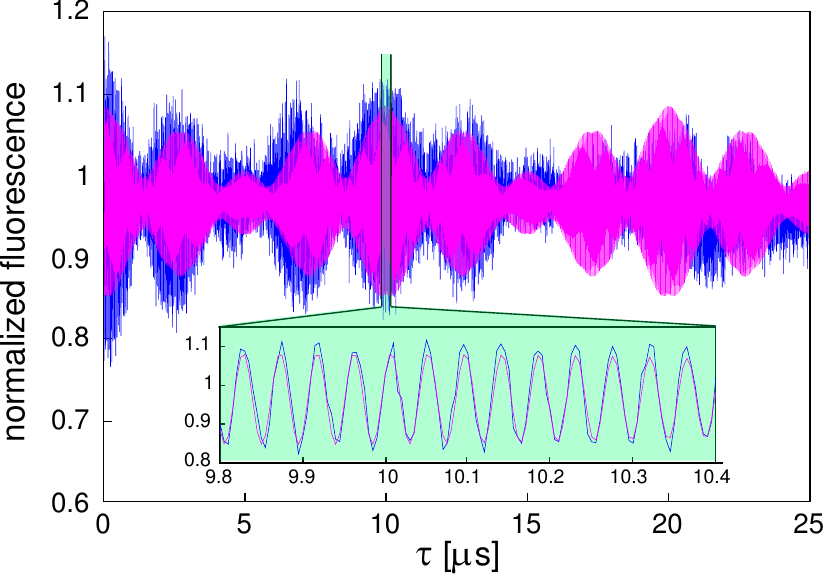}
  \caption{Electron spin Rabi oscillations. Solid blue line: experimental data,
  solid magenta line: result from Fourier transform (Fig.\ref{fig:fft}).
  Three beating cosine with frequencies corresponding to the three local maxima
  in the FFT are plotted.}
  \label{fig:rabi}
\end{figure}

\begin{figure}
  \centering
  \includegraphics{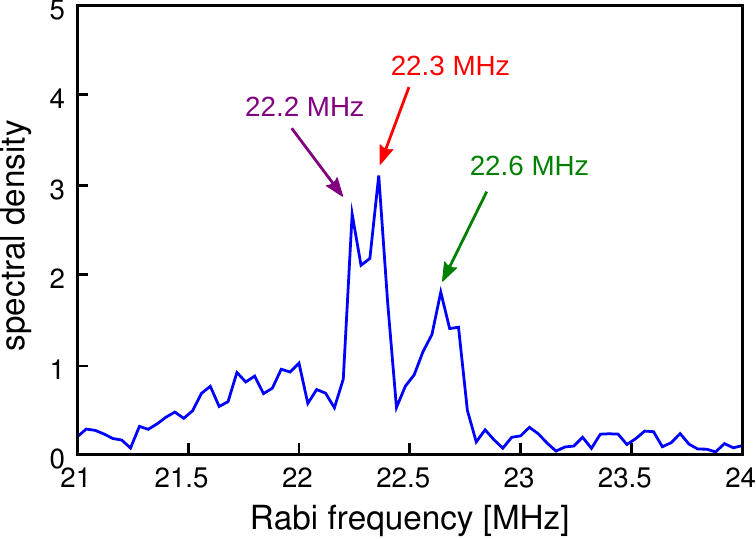}
  \caption{Fourier transform of the Rabi signal. The data shows three
  distinct Rabi frequencies corresponding to the individual hyperfine levels.}
  \label{fig:fft}
\end{figure}

To measure Rabi beats, a coplanar waveguide (gap=$10\mu$s, width=$10\mu$s) was
fabricated directly onto a type IIa diamond sample with $\langle 100\rangle$
crystallographic orientation and less than 4ppb nitrogen concentration
(element6, electronic grade). A DCmagnetic field was applied with a permanent
magnet. A microwave field
resonant to the lowest of the hyperfine transitions (2.4524GHz) was generated
with a synthesizer (Rhode\&Schwarz SMIQ) and applied to the sample through a
fast switch (minicircuits ZASW-2-50DR). A single NV center was identified in
the coplanar gap using a home built confocal fluorescence microscope.
Rabi oscillations were measured using a standard laser- and microwave pulse
sequence \cite{jelezko2004observation}. Figure \ref{fig:rabi} shows the corresponding Rabi oscillations. The data shows a base oscillation of about $42$MHz (see inset). Due to the beating this is approximately twice the base
Rabi frequency. The two beat frequencies $\delta\Omega_{0}=100$kHz and
$\delta\Omega_{+1}=400$kHz correspond to the $m_s=0$ and $m_s=+1$ hyperfine
transition. Thus in total three Rabi oscillations are observed. Their individual
frequencies are seen in the Fourier transform (Figure \ref{fig:fft}),
that shows $\Omega_{-1}=22.2$MHz, $\Omega_0=22.3$MHz, and
$\Omega_{+1}=22.6 $MHz. Note that the FFT data is less pronounced due to the
relatively sparse sampling of less then 10 points per period (see also zoom in
Fig. \ref{fig:rabi}) and power drift during the measurement. The sparse
sampling was necessary to keep the total measurement time (several days for Fig. \ref{fig:rabi}) within reasonable
bounds. To complete our measurment of the hyperfine splitting from the Rabi
beats, we use Eq.\ref{eq:detuning} and convert the measured beats into energy
level shifts. We find $\delta_{0}=2.1$MHz and $\delta_{-1}=4.2$MHz, which is in
good agreement with the ESR spectrum (Fig.\ref{fig:esr}).

The Rabi oscillations show a decay $T_1^*\approx 25\mu$s. This decay is much
much smaller than the expected $T_1^\rho\approx 1$ms and is limited by
microwave power drift as we show below. With the present decay $T_1^*$, the
resolution of the hyperfine measurement is as follows. The resolution of a
hyperfine measurement from the Rabi beats is determined by the Rabi frequency
and the number $N$ of visible Rabi oscillations.  To measure a small change in
the Rabi period, we need to measure a number  of oscillations. Combining
Eq.(\ref{eq:N}) with Eq.(\ref{eq:detuning}), we  have
\begin{equation}
\delta=\frac{\Omega}{\sqrt{N}}=\frac{2\pi}{\sqrt{TT_1^*}}.
\end{equation}
In our case we find $\delta\approx 6$MHz, which is in good qualitative
agreement with our observations, since our data, in particular the Fourier
transform, Fig.\ref{fig:fft}, implies that we can indeed resolve the 4.32MHz detuning
related to the farther detuned hyperfine level (22.6MHz peak in the FFT),
however, we can just barely resolve the 2.16MHz detuning related to the lesser
detuned hyperfine level (22.3MHz peak in the FFT). In the present measurements the
total number of Rabi oscillations is about 500, which translates into an
equivalent spatial resolution of about 10nm. We expect that an improvement of
the power stability to a level better than $10^{-4}$ should be achievable
without a large technical effort, and that spatial drifts of the microwave field
should still be negligable in this case. Thus a spatial resolution of about 1nm
should be achievable in a realistic measurement.

\begin{figure}
  \centering
  \includegraphics{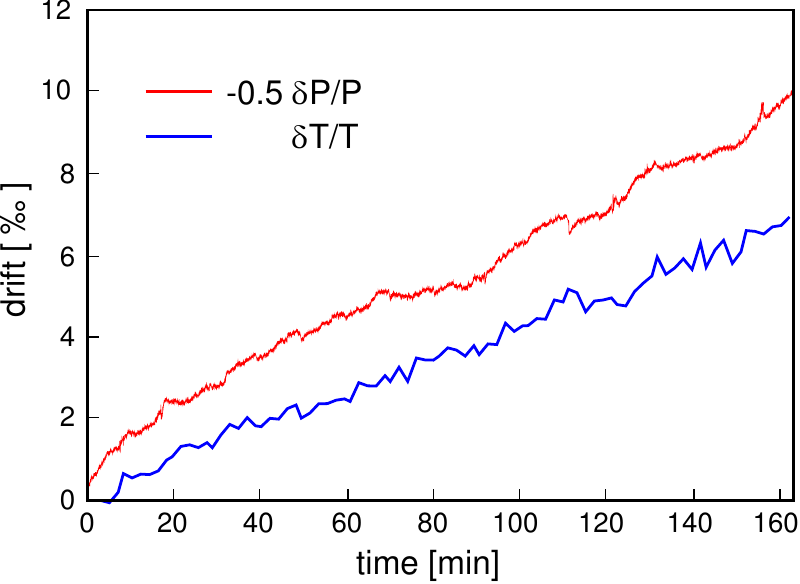}
  \caption{Power drift. The relative change of the Rabi period is compared to
  the relative change of the microwave power.}
  \label{fig:power}
\end{figure}

Any fluctuation or drift of the applied microwave power result in a
small shift of the Rabi frequency over time, that washes out the oscillations.
To proof that the observed decay is caused by power drift, we
monitor the power transmitted through the sample and compare it to the Rabi period. This measurement is realized
by successively performing short Rabi measurements in a small window ranging
from $25.0\mu$s to $25.1\mu$s. The result is shown in Figure \ref{fig:power}.
Since the Rabi period is inversely proportional to the frequency and the  frequency is proportional to the square
root of the power, we can express the  relation between the relative change of
the Rabi period as
\begin{equation}
\frac{\delta T}{T}=-\frac{1}{2}\frac{\delta P}{P}.
\end{equation}
The data confirms that the decay time of the Rabi
oscillations is limited by drift of the microwave power, which is slightly better than
$10^{-3}$ over 24h in our case. Note that we also observed a change of the
transmitted microwave power when the microscope objective was moved (about $10^{-3}/\mu$m). This is explained by a
change of the parasitic capacitance of the coplanar waveguide, which plays a
role owing to the close proximity of the high N.A. microscope objective to the
sample (working distance $200\mu$m).

With the fast Rabi frequency required for a high spatial resolution, it is no
longer feasible to shift the $m_s=\pm 1$ spin levels sufficiently far apart from
each other such that a single spin transitions is driven selectively.
This is because the separation between the $m_s=\pm 1$ levels needed to be
larger than the Rabi frequency. This would in turn require a large
magnetic field. However at large magnetic field, level mixing
causes the optical ESR contrast to vanish, unless the magnetic field
is aligned parallel to the NV axis \cite{neumann2010single}. Alignment of the
magnetic field parallel to the NV axis is however not possible in a sample with many randomly
distributed NV centers. Thus, in a high resolution measurement, we will always
measure with a small DC bias field and hence we will simulstaneously drive the
$m_s=\pm 1$ spin transitions. In this case, six hyperfine levels are present
whose Rabi oscillations beat with each other. This situation
differs from the previous experiments. While in the previous case, the Rabi
signal was an incoherent sum of three oscillations (during a measurement, the
nuclear spin switched randomly and slowly between the different spin states),
now each nuclear spin manifold forms a V-type energy level
scheme that is driven simultaneously. In the following, we evaluate the
corresponding Rabi beat signal.

The NV's spin Hamiltonian with external magnetic field reads
\begin{equation}
 H=\op{H}_\text{ZFS}+\op{H}_\text{Z}
\end{equation}
with
\begin{align}
 \op{H}_\text{ZFS}&=D\,(\op{S}_z^2-\frac{2}{3}\,\eins)+E\,(\op{S}_x^2-\op{S}_y^2)\\
 \op{H}_\text{Z}&=g\,\beta\,B\,\op{S}_z.
\end{align}
After transforming into the energy Eigenbasis where we name the $m_s=0$ state
as $\ket{0}$ and the two state that correspond to the $m_s=\pm 1$ states as
$\ket{1}$ and $\ket{2}$. Now the semiclassical microwave field is applied
\begin{equation}
 \op{H}_\text{MW}=\lambda\,\exp{\i\,\omega\,t}\,\op{S}_x.
\end{equation}
here we assume, that both transitions have the same transition matrix elements,
which breaks down with rising $E$. Now do the rotating wave approximation (RWA)
and transform into the rotating frame which leaves us with a Hamiltonian of the
form
\begin{equation}
 \op{H}_\text{rot}=\left(\begin{array}{ccc}
                    0 & \lambda & \lambda\\
                    \lambda & \Delta-\delta & 0\\
                    \lambda & 0 & \Delta+\delta
                   \end{array}\right).
\end{equation}
where $\delta=\Delta E_{\ket{1}\,\ket{2}}/2$ and $\Delta$ the detuning of the
microwave from $\ket{1}+\delta$.

To obtain an analytical solution we set $\Delta=0$. Calculation of the
Eigenvalues leads to the Rabi frequency $\sqrt{2\,V^2+\delta^2}$ which, were
the two levels detuned is $\sqrt{2}$ times the Rabi frequency of driving a
single level. We now take a look at the time evolution of the system with
initial state $\ket{0}$. The dynamics of $\ket{0}$ are
\begin{equation}
 \rho_{\ket{0}}=\frac{\left(\delta^2+2\,\lambda^2\,\cos{\left[\sqrt{2\,\lambda^2+\delta^2}\,t\right]}\right)^2}{(2\,\lambda^2+\delta^2)^2}.
\end{equation}
from here we see, that there are actually two oscillation frequencies involved,
the Rabi frequency and double the Rabi frequency corresponding to the $\cos{}^2$
term. The latter will be the base Rabi frequency observed in experiments, and we
find the identity for the Rabi beat as
\begin{equation}
\delta\Omega=\frac{2\delta^2}{\Omega_0},
\end{equation}\label{eq:DetuningV}
which is different from the previous expression, Eq.(\ref{eq:detuning}).

\begin{figure}
  \centering
  \includegraphics{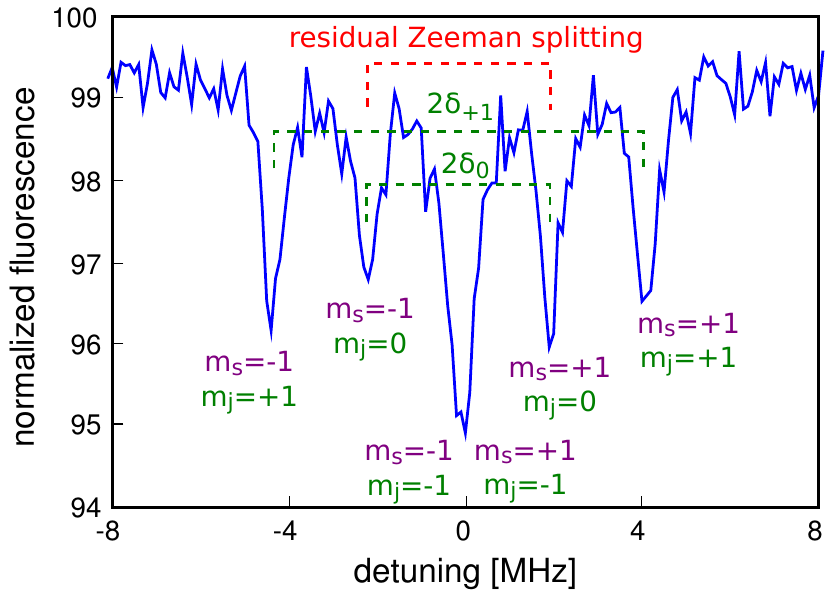}
  \caption{ESR spectrum of the second NV center without DC bias field. The two
  hyperfine triplets corresponding to the $m_s=-1$ and $m_s=+1$ spin levels
  overlap in the central dip. Detuning is denoted relative to the central
  hyperfine transition corresponding to 2.8706GHz.}
  \label{fig:DegenerateESR}
\end{figure}

To measure Rabi beats with V-type microwave excitation, we remove the DC
bias field. Moreover we pick a different NV center that is closer to the
central conductor and feels a higher microwave power. Figure \ref{fig:DegenerateESR}
shows the ESR spectrum, with a total of five dips. The central dip is about
twice larger as compare to the other dips. In this experiment, the Zeeman
splitting due to a residual magnetic field (earth magnetic field as well as
magnetized parts of the setup) closely matches the hyperfine splitting. As a
consequence, the two hyperfine triplets corresponding to the $m_s=-1$ and 
$m_s=+1$ spin transition are shifted in such a way that the highest peak
of the lower triplet overlaps  with the lowest peak of the higher triplet,
resulting in five dips with a pronounced central dip. Rabi oscillations are now
driven with the microwave frequency tuned in resonance with the central  dip
corresponding to 2.8706GHz. This case corresponds to the theoretical case
$\Delta=0$ treated above, with $\delta_{0}=2.18$MHz and
$\delta_{+1}=4.36$MHz for the driving of the $m_j=0$ and $m_j=+1$
manifold, respectively.

\begin{figure}
  \centering
  \includegraphics{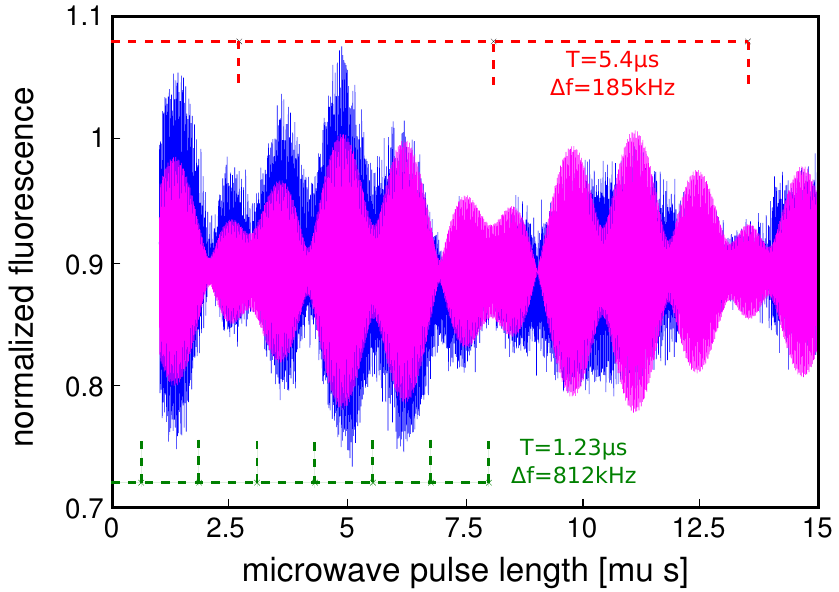}
  \caption{Rabi oscillations under V-type microwave excitation. Solid blue
  line: experimental data. Red and green markers denote the extracted beat
  frequencies. Solid magenta line: result of the extracted beat frequencies.
  Three beating cosine with the corresponding frequency shifts are plotted.}
  \label{fig:DegenerateRabi}
\end{figure}

\begin{figure}
  \centering
  \includegraphics{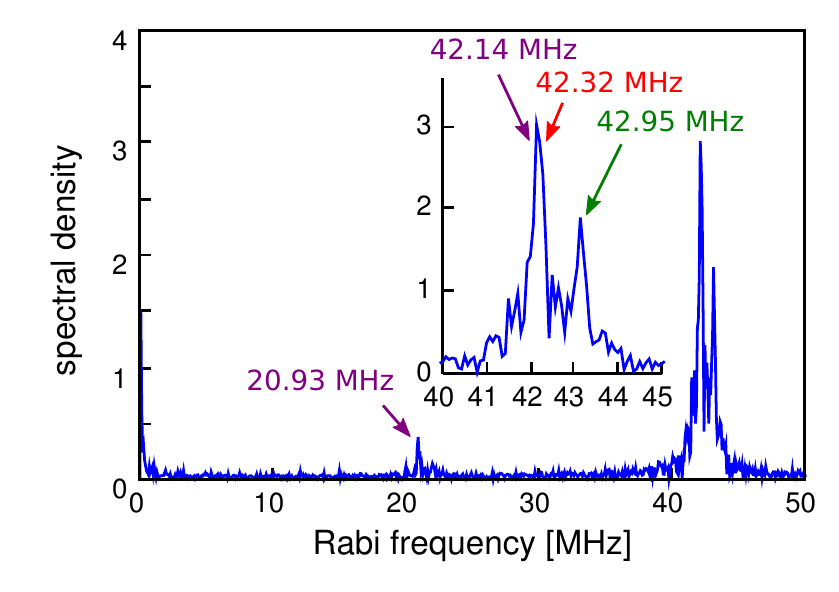}
  \caption{Fourier transform of the Rabi oscillations under V-type
  microwave excitation. Peaks around 42MHz and 21MHz correspond to the base Rabi
  oscillations and a modulation at half the frequency. The inset shows 
  the beating Rabi frequencies around 42MHz. Labels mark the three
  frequencies as extracted from the beat analysis of the raw data. The FFT
  resolves the fast beat, while it does not resolve the slow beat due to sparse
  sampling.}
  \label{fig:DegenerateFFT}
\end{figure}

The Rabi oscillations and their Fourier transform are shown in Figure
\ref{fig:DegenerateRabi} and \ref{fig:DegenerateFFT}. The first
observation is that the Rabi frequency is increased by
about a factor of two as compared to the previous case. Two effects contribute
here. First, with lambda type driving the Rabi frequency is
enhanced by a factor of $\sqrt{2}$ as compared to driving a single transition,
and second, the Rabi frequency is additionally increased due to the higher
microwave power in the close vicinity of the central conductor. The second
observation is a fast and a slow beat with frequency of about 185kHz and
812kHz, respectively. These beats are seen most clearly in the Rabi
oscillations. Note that the slower beat is not resolved in the FFT
(Fig.\ref{fig:DegenerateFFT}) due to power drift and sparse sampling. We can
now convert the observed beatings to energy level shifts according to
Eq.(\ref{eq:DetuningV}). We find $\delta_{0}=2.0$MHz and
$\delta_{+1}=4.1$MHz, which is in good agreement with the ESR spectrum.

\section{Summary and Conclusions}

We have performed two proof-of-principle experiments towards $T_1$ limited
magnetic resonance imaging with NV centers in diamond. First, we have
demonstrated the measurement of a large number ($>500$) of Rabi flops, and we
have shown that the hyperfine interaction due to $^{14}$N can be resolved from
such a measurement. Second, we have studied the Rabi beats without a large DC
bias field, where the nuclear spin manifolds form V-type energy level schemes.
The base Rabi frequency is increased by $\sqrt{2}$ and the time evolution of the
state population is modulated by an oscillation with half the base Rabi
frequency. While the present experiments were performed in bulk, in the future,
it will be an important step to demonstrate Rabi beat imaging with NV centers
embedded in diamond nano crystals. The present experiments are relevant to
quantum information processing in diamond. The ability to drive a large number of
Rabi flops could allow to precisely control the spin state of several (detuned)
NV centers simultaneously. We envision a microwave pulse with
precisely adjusted length that performs independent unitary
transformations on a quantum register, such as rotate an NV A to $|1\rangle$
while simultaneously rotating an NV B to the superposition
$\sqrt{2}(|0\rangle+|1\rangle$.

\section{Acknowledgements}
This work was supported by the European Union, Deutsche Forschungsgemeinschaft
(SFB/TR21 and FOR1482), Bundesministerium f\"ur Bildung und Forschung, and the
Landesstiftung Bandenw\"urttemberg.


\end{document}